\documentclass[10pt]{iopart}
\usepackage{perpage} 
\usepackage{graphicx}
\usepackage[usenames]{color}

\newcommand{\IncludeFig}[1]{\includegraphics[width=0.48\textwidth]{#1}}
\newcommand{\Header}[1]{\section{#1}}
\newcommand{\rms}{r_{nn}^{\rm{rms}}}
\def\lpc{$^1$}\def\LPC{LPC Caen, Normandie Universit\'e, ENSICAEN, Universit\'e de Caen, CNRS/IN2P3, Caen, France}
\def\lln{$^2$}\def\LLN{CRC/UCL, Louvain-la-Neuve, Belgium}
\def\bam{$^3$}\def\BAM{School of Physics and Astronomy, University of Birmingham, Birmingham B15~2TT, UK}
\def\mad{$^4$}\def\MAD{Instituto de Estructura de la Materia, CSIC, E-28006 Madrid, Spain}
\def\ulb{$^5$}\def\ULB{Universit\'e Libre de Bruxelles, CP 226, B-1050 Bruxelles, Belgium}
\def\sur{$^6$}\def\SUR{Department of Physics, University of Surrey, Guildford, Surrey, GU2~7XH, UK}
\def\sco{$^7$}\def\SCO{Department of Engineering and Science, University of the West of Scotland, Paisley PA1 2BE, UK}
\def\iso{$^8$}\def\ISO{ISOLDE, CERN, CH-1211 Geneva 23, Switzerland}
\def\ire{$^9$}\def\IRE{Institut de Recherche Subatomique, IN2P3-CNRS, Universit\'e Louis Pasteur, BP 28, F-67037 Strasbourg cedex, France}
\usepackage{iopams}  

\begin{document}

\title[Chronology of the three-body dissociation of $^8$He]{Chronology of the three-body dissociation of $^8$He}

\author{B.~Laurent\lpc\footnote{Present address: CEA, DAM, DIF, F-91297 Arpajon, France.},
 F.M.~Marqu\'es\lpc, C.~Angulo\lln\footnote{Present address: SCK-CEN, Boeretang 200, 2400 Mol, Belgium.},
 N.I.~Ashwood\bam, M.J.G.~Borge\mad, V.~Bouchat\ulb,
 W.N.~Catford\sur, N.M.~Clarke\bam, N.~Curtis\bam, M.~Freer\bam, F.~Hanappe\ulb, V.~Kinnard\ulb,
 M.~Labiche\sco\footnote{Present address: STFC Daresbury Laboratory, Daresbury, Warrington WA4 4AD, UK.},
 T.~Materna\ulb, P.~McEwan\bam, T.~Nilsson\iso, A.~Ninane\lln,
 G.~Normand\lpc\footnote{Present address: GANIL, Caen, France.},
 N.A.~Orr\lpc, S.D.~Pain\sur, E.~Prokhorova\ulb, L.~Stuttg\'e\ire, C.~Timis\sur
}
 
\address{\lpc\LPC}
\address{\lln\LLN}\address{\bam\BAM}\address{\mad\MAD}\address{\ulb\ULB}\address{\sur\SUR}\address{\sco\SCO}\address{\iso\ISO}\address{\ire\IRE}
\ead{marques@lpccaen.in2p3.fr}
\vspace{10pt}
\begin{indented}
\item[]\today
\end{indented}

\begin{abstract}
 The space and time configurations of the dissociation of $^8$He into $^6$He+$n$+$n$, on C and Pb targets,
 have been explored simultaneously for the first time. The final-state interactions in the $n$-$n$ and $^6$He-$n$
 channels are successfully described within a model that considers independent emission of neutrons from a Gaussian
 volume with a given lifetime. The dissociation on C target exhibits a dominant sequential decay through the ground
 state of $^7$He, consistent with neutrons being emitted from a Gaussian volume of $\rms=7.3\pm0.6$~fm with a
 $n$-$n$ delay in the sequential channel of $1400\pm400$~fm/$c$, in agreement with the lifetime of $^7$He.
 The lower-statistics data on Pb target correspond mainly to direct breakup, and are well described using the
 $n$-$n$ volume measured, without any $n$-$n$ delay. The validity of the phenomenological model used is discussed.
\end{abstract}

%


%
\ioptwocol

\Header{Introduction}
 Along the neutron dripline some nuclear systems develop an extreme clustering structure already in their ground
 state, in which the weakly bound neutrons form a halo around the rest of the system, or core \cite{Jen04}.
 The most exciting of these are the Borromean two-neutron halo systems, $^6$He, $^{11}$Li and $^{14}$Be,
 exhibiting a bound core-$n$-$n$ structure where all the two-body subsystems are unbound \cite{Zhu93}. As such,
 these systems are unique for the study of three-body correlations. 
 Other heavier candidates, $^{17,19}$B and $^{22}$C, have not been explored in detail yet, and the structure of
 $^8$He seems to be dominated by an $\alpha$ core plus four neutrons \cite{Mue07}.
 
 The breakup of these genuine three-body systems unveils a complex interplay between the nature of their decay,
 either direct or sequential \cite{Ego12,Jon15}, and the resonances and final-state interactions (FSI) in the
 two-body channels. Compared to systems involving charged particles, two-neutron haloes convey a clearer picture
 of the decay due to the absence of Coulomb forces. The population of core-$n$ resonances dictates the
 sequential nature of the decay, and the $n$-$n$ interaction modifies their final state at low relative momentum
 \cite{Wat52,Mig55,Led82}. In particular, Ref.~\cite{Led82} links this modification with the space-time proximity
 of the neutrons, under the hypothesis of independent emission from a Gaussian source, by exploiting the principle
 that the effect of the short-range nuclear force will be stronger at shorter distances.
 
 The dissociation in the field of a heavy target can be used at intermediate energies to induce direct breakup
 of two-neutron haloes \cite{Iek93,Nak06}. In that case, no core-$n$ resonances are populated and,
 according to Ref.~\cite{Led82}, the only parameter characterizing the decay should be the space dimension of the
 neutron emission. This approach was used to estimate the average $n$-$n$ separation at breakup
 in $^6$He, $^{11}$Li and $^{14}$Be, leading to $\rms$ values of 6--7~fm
 in agreement with predictions from three-body models \cite{FMM00}.
 However, in the case of $^{14}$Be it was found that breakup on a lighter target introduced a sequential channel
 through various resonances in $^{13}$Be \cite{FMM01}, that induced an average delay between the emission of the
 neutrons and thus reduced the $n$-$n$ signal. The delay, $150^{+250}_{-150}$~fm/$c$, was attributed to a series
 of resonances in the $^{13}$Be system, their number and energy being not completely resolved.

 A similar decrease of the FSI effects in a nucleon pair due to sequential decay was recently observed in
 two very different scenarios, the two-neutron decay from the continuum of $^{18}$C and $^{20}$O \cite{Rev18}
 and the two-proton decay of unbound states of $^{22}$Mg and $^{23}$Al \cite{Fan16}. In both works, the direct
 and sequential channels were clearly identified, and the analysis using the simple FSI model with a space-time
 Gaussian source lead to source sizes of the order of a nucleus of mass 20. The delays obtained,
 however, could not be assigned to individual intermediate resonances and, as in Ref.~\cite{FMM01},
 were only considered qualitatively.
 
 The possibility to measure a neutron delay generated by the well-defined lifetime of an intermediate
 core-$n$ resonance represents a unique opportunity to map the space-time decay of a three-body system.
 Ideally, the core-$n$ subsystem should have one state and if possible with narrow width
 (long lifetime), that would induce a significant delay between the emission of the neutrons.
 While $^8$He is understood as a four-neutron halo or skin \cite{Mue07}, its three-body $^6$He+$n$+$n$
 breakup channel exhibits the characteristics noted above: only one state is known in $^7$He decaying into
 $^6$He+$n$, and it is narrow enough ($\Gamma=150\pm20$~keV \cite{He7}) to significantly delay the emission
 of the second neutron ($\hbar/\Gamma=1320\pm180$~fm/$c$).

 In this Letter, we report on the dissociation of a $^8$He beam into $^6$He+$n$+$n$ on C and Pb targets. The
 formation of $^7$He has been clearly identified in Dalitz plots, showing that the breakup on C target is mainly
 sequential and goes through the ground state of $^7$He. The description of the three-body final state
 with the FSI model used in Refs.~\cite{FMM01,Rev18}
 leads for the first time to the simultaneous measurement of the $n$-$n$ volume at breakup and the delay between
 the emission of the neutrons, the latter being in very good agreement with the lifetime of $^7$He.
 These observations are confirmed in the breakup on Pb target, in which no $^7$He is observed and that is
 successfully described by the direct emission of the neutrons from the same volume and without any delay.
 
\Header{Experiment}
 A $^8$He secondary beam of 15~MeV/nucleon and 10$^4$~pps was produced at GANIL using SPIRAL, and was
 tracked onto the breakup targets (95~mg/cm$^2$ C and 284~mg/cm$^2$ Pb) on a particle-by-particle basis using
 a thin plastic detector (100~$\mu$m of BC408) and a drift chamber at 11~cm upstream of the target. The $^6$He
 fragments were identified using two 500~$\mu$m Si strip detectors and sixteen 2.5~cm CsI crystals from the
 CHARISSA collaboration inside the reaction chamber, and the neutrons detected using 90 elements of the DEMON
 array, placed in a staggered arrangement at forward angles in order to minimize the contribution of cross-talk
 (the scattering of one neutron through several detectors) \cite{fmm00}.

\begin{figure}[t]
\begin{center}
 \IncludeFig{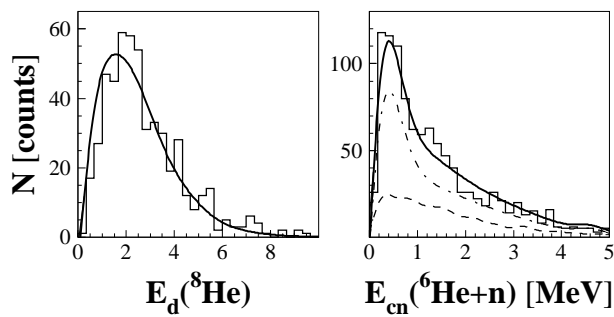}
 \caption{Reconstructed decay energy of the $^6$He+$n$+$n$ system (left) and the $^6$He+$n$ subsystem (right),
 for the breakup of $^8$He on a C target. The solid lines correspond to the input (left) and result (right) of
 the simulation described in the text, decomposed on the right panel into its sequential decay (dot-dashed)
 and direct breakup (dashed) components.} \label{f:Ecn}
\end{center}
\end{figure}

 With the momenta of the core+$n$+$n$ breakup fragments, we calculate the two and three-body invariant masses
 $M_{cn}$ and $M_{cnn}$, and then the decay energy of the system $E_d=M_{cnn}-m_c-2m_n$
 and the relative core-$n$ energy $E_{cn}=M_{cn}-m_c-m_n$ (Fig.~\ref{f:Ecn}).
 Although $E_d$ peaks at about 2~MeV, the energy of a resonance observed in a previous work \cite{Mak01},
 we have not assumed the population of individual resonances due to the limited acceptance (that decreases
 beyond 2~MeV) and resolution (about 1.5~MeV {\sc fwhm} at 2~MeV), and have focused on the final-state
 correlations emerging from this continuum energy distribution.

 The interacting phase-space model used  was introduced in Ref.~\cite{FMM01}. In brief, the experimental $E_d$
 distribution is used as input to generate events following three-body or twice two-body phase space \cite{Nik68},
 respectively for direct or sequential decay. The latter is described as the breakup into one neutron
 and a core-$n$ resonance, with a relative energy $E_{cn}$ given by a Breit-Wigner line-shape with energy-dependent
 width of parameters $(E,\Gamma)$, which is then allowed to decay into the core plus the second neutron.
 The $n$-$n$ FSI is introduced via a probability to accept the event as a function of the $n$-$n$ relative
 momentum $q_{nn}=|\vec{p}_{n_1}-\vec{p}_{n_2}|$ following the form of the $n$-$n$ correlation function
 $C_{nn}$ \cite{Led82,FMM00}, defined by the neutron emission volume.
 The final momenta are filtered through a simulation, including all experimental effects \cite{FMM00,fmm00},
 and the different observables are reconstructed.
 
\Header{The decay model}
 In the analog case of the decay of core-$p$-$p$ systems, the correlations in the $p$-$p$ channel have been often
 interpreted in terms of the microscopic structure of the two-proton emitter, like for example in the $2p$
 decay of $^6$Be and $^{45}$Fe \cite{Gri09}. Within this formalism, the relative $p$-$p$ energy is given by
 the configuration mixing of the three-body wave function of the emitter, that may lead to an `anti-correlation'
 (strength at high relative energy, back-to-back emission) or even to oscillatory patterns \cite{Gri09,Gri09b}.
 Recently, a similar formalism has been proposed for core-$n$-$n$ decays \cite{Gri18}.

\begin{figure}[t]
\begin{center}
 \includegraphics[width=0.44\textwidth]{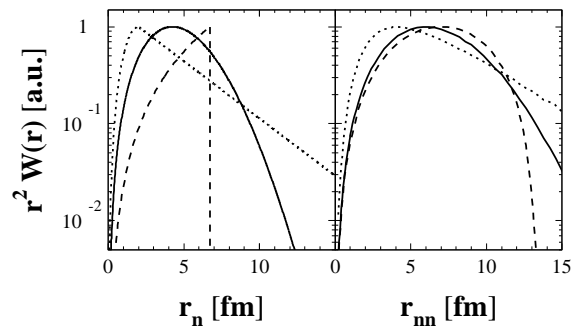}
 \caption{Density distributions (left) corresponding to a Gaussian of sigma 3~fm (solid), a sphere (dashed),
 and a Yukawa (dotted line), all with the same rms radius. The distributions of relative distance between two
 independent particles within these volumes are shown on the right.} \label{f:Wxs}
\end{center}
\end{figure}
 
 However, the core-$n$-$n$ final states measured to date do not exhibit any such patterns. In the $2n$ decay
 from the continuum of $^5$H \cite{Nor04,Gol05}, $^6$He \cite{FMM00,Nor04}, $^8$He (this work), $^{11}$Li
 \cite{FMM00,Smi16}, $^{14}$Be \cite{FMM00,FMM01}, $^{16}$Be \cite{Spy12}, $^{18}$C and $^{20}$O \cite{Rev18},
 the strength is systematically located at low $n$-$n$ energy, as predicted by Refs.~\cite{Wat52,Mig55}.
 This low-energy signal is originated by the $s$-wave $n$-$n$ interaction, and is in fact used in the reaction
 with the simplest two-neutron final state, $^2$H$(\pi^-\!,\gamma)2n$, to determine the $n$-$n$ scattering length
 $a_{nn}$ \cite{Gab84,Pre75}. In that system, the geometry of the process is well known (given by the deuteron
 wave function) and the only free parameter in the description of the low $q_{nn}$ signal (or its counterpart at
 high photon energy) is $a_{nn}$.
 
 Ref.~\cite{Led82} extends the formalism of Refs.~\cite{Wat52,Mig55} by taking into account the influence
 of the two-nucleon proximity on the effects of their interaction (see appendix for the relevant formulas).
 It considers the effect of the $s$-wave scattering amplitude, dominant at low energies, on a pair of nucleons
 separated by a four-momentum distance, and integrates it over a source in space and time.
 The resulting factor becomes a probability distribution that modulates phase space, as in Ref.~\cite{Pre75}.
 This model described accurately the low-energy peaks observed in
 the $n$-$n$ final state of previous works \cite{FMM00,FMM01,Rev18,Nor04,Smi16,Spy12}, although contrary to
 Refs.~\cite{Gab84,Pre75} the value of $a_{nn}$ was fixed and it was the size of the source that was varied.

 The formalism of Ref.~\cite{Led82} was, however, developed for a {\it Gaussian\/} source emitting
 {\it independent\/} neutrons. Obviously, we do not pretend that the wave function of $^8$He is Gaussian, nor
 the two valence neutrons independent. First, the fact that the neutrons move independently in a Gaussian source
 was in part exploited in Ref.~\cite{Led82} for analytical ease, as in that case the distribution of relative
 distance is also Gaussian. 
 However, $C_{nn}$ does not depend on each neutron's position in the source but on their relative distance,
 so we can directly choose the shape of the latter as input of the model, without any hypothesis on the overall
 matter distribution of $^8$He.
 Moreover, the distributions of relative distance $W(r_{nn})$ in general soften the shape of the individual
 distributions $W(r_n)$, as can be seen in the examples of Fig.~\ref{f:Wxs} for three very different shapes.ç
 This may explain in part why the Gaussian hypothesis has been able to describe reasonably well
 the final states of previous works, even for systems that were not supposed to be Gaussian.

 A second approximation of Ref.~\cite{Led82} is the neglect of internal momentum correlations of the form
 $W(r_{nn},q_{nn})$. We will assume that those potential correlations are small, or that they have a negligible
 impact on the correlation factor after averaging over the whole source. The ability of the model to describe,
 at least at first order, the very specific channel subject of this work represents a severe test that will
 confirm or refute the validity of these two approximations.

\Header{Sequential decay and time}
 In Ref.~\cite{FMM00}, the analysis of the dissociation of halo nuclei on a Pb target assumed simultaneous
 emission of both neutrons in the Coulomb field of the target. When there is no emission delay, the correlation
 function of a Gaussian source becomes analytical (\ref{e:Cnn0}). If the dissociation is sequential,
 however, the emission of the neutrons cannot be considered simultaneous and a space-time analysis is needed.
 The effect of the $n$-$n$ FSI depends then on the two space and time parameters ($r_0,\tau_0$), that correspond
 to the sigma values of the Gaussian space-time source (\ref{e:B0},\ref{e:Bi}).
 
 In this work, the $^8$He projectile is considered to be excited by the C/Pb target into the continuum, and
 the unbound $^6$He+$n$+$n$ system decays in flight, either directly into the three particles, or through an
 intermediate $^6$He+$n$ resonance. Therefore, the only free parameters of our interacting phase-space model
 are three: $\rms$, the root-mean-square $n$-$n$ distance ($\sqrt{6}\,r_0$);
 $P(^7\rm{He})$, the probability of sequential decay through $^7$He ground state;
 and $\tau$, the neutron delay introduced by this $^7$He resonance ($\sqrt{2}\,\tau_0$).
 
\begin{figure}[t]
\begin{center}
 \IncludeFig{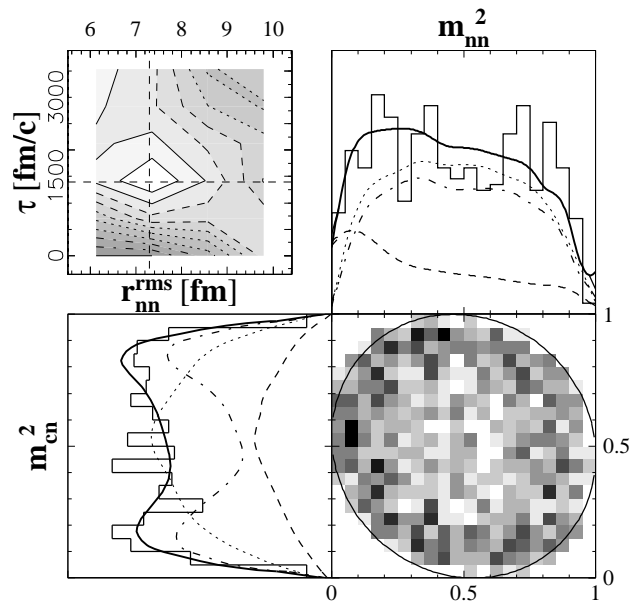}
 \caption{Dalitz plots (core-$n$ versus $n$-$n$), and the projections onto both axes, for the data from the
 dissociation of $^8$He into $^6$He+$n$+$n$ on a C target. The lines correspond to the best fit of the simulation
 (solid), with the contributions of sequential (dot-dashed) and direct (dashed) breakup. The dotted line
 corresponds to phase space. The inset shows the $\chi^2$ between simulations and data as a function of the delay
 between the neutrons and their average distance at breakup (each contour corresponds to one unit of $\chi^2$).} \label{f:dal}
\end{center}
\end{figure}
 
 The Dalitz plot of the decay after dissociation on the C target is shown in Fig.~\ref{f:dal}, as a function of
 the $n$-$n$ and $^6$He-$n$ invariant masses normalized between 0--1 to the available energy $E_d$ \cite{FMM01}.
 In the absence of interactions/correlations, the whole plot should be populated uniformly, and the projections
 should follow the phase-space dotted lines. On the other hand, the attractive $n$-$n$ interaction would
 overpopulate the low $m_{nn}^2$ part, and resonances due to the $^6$He-$n$ interaction would lead to horizontal
 bands \cite{FMM01}.

 The Dalitz plot exhibits a clear crescent shape as a result of both interactions, a slight increase at low
 $m_{nn}^2$ and two horizontal bands at $m_{cn}^2\sim0.15$ and 0.85, leading to a depletion at the center.
 This is more easily observed in the projections shown on the same figure, where the slight increase
 towards $m_{nn}^2=0$ and the two peaks on the wings of the $m_{cn}^2$ distribution become more evident with
 respect to the expected phase-space distribution (dotted line). Both interactions are clearly present since none
 of them (dot-dashed or dashed lines) is able to reproduce the data on its own.

\begin{figure}[t]
\begin{center}
 \IncludeFig{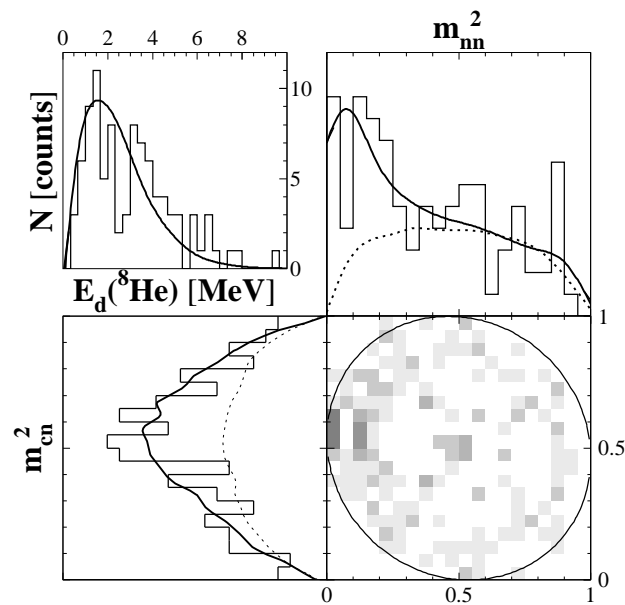}
 \caption{The same Dalitz plot shown in Fig.~\ref{f:dal} but for the dissociation of $^8$He into $^6$He+$n$+$n$
 on a Pb target. The solid line corresponds to the simulation of direct breakup with $\rms=7.3$~fm, and the dotted
 line to phase space. The inset shows the reconstructed decay energy of the $^6$He+$n$+$n$ system, with the solid
 line being the input of the simulation.} \label{f:mc2}
\end{center}
\end{figure}

 The probability of sequential decay can be easily extracted from the already mentioned characteristic
 signals of $^7$He ground state in the $^6$He+$n$ distributions, the narrow peak in $E_{cn}$ (Fig.~\ref{f:Ecn})
 and the symmetric wings in $m_{cn}^2$ (Fig.~\ref{f:dal}). A fit of the latter gave a value of
 $P(^7\rm{He})=70^{+10}_{-5}$\% for the data on C target.
 Using this result, we have combined the direct and sequential scenarios and varied the two parameters
 $(\rms,\tau)$ describing the space-time characteristics of the neutron source.
 The determination of the values that describe best the $n$-$n$ signal in $m_{nn}^2$ has been undertaken
 through the calculation of the $\chi^2$ between each simulation and the experimental distribution,
 leading to the $\chi^2$ surface shown on the inset of Fig.~\ref{f:dal}. A clear minimum appears at
 $\rms=7.3\pm0.6$~fm and $\tau=1400\pm400$~fm/$c$. The simulation corresponding to this minimum is represented
 by the solid line in the projections of Fig.~\ref{f:dal}.
 Note that the $n$-$n$ FSI acts on all events, depending on the space dimension for direct breakup ($\rms$),
 and on both the space and time dimensions for sequential breakup ($\rms,\tau$).

 The measured delay between the emission of the neutrons corresponds well to the expected scenario, the
 lifetime of $^7$He ($1320\pm180$~fm/$c$). Regarding the size of the $n$-$n$ volume, it should be considered
 as an average value for the $^8$He continuum up to 5~MeV beyond the $^6$He+$n$+$n$ threshold.
 We note that this analysis is similar to the one used in Refs.~\cite{Fan16,Col95}, in which the $\chi^2$
 between the experimental and theoretical correlation functions was minimized in order to extract the size and
 lifetime of a $p$-$p$ emitter \cite{Fan16} and of a compound nucleus evaporating neutrons \cite{Col95}.

\Header{Direct decay and space}
 The dissociation on a Pb target lead to lower statistics, that prevented the Dalitz plot analysis or the
 construction of the $\chi^2$ surface.
 Nevertheless, in Fig.~\ref{f:mc2} we show the plot and its invariant-mass
 projections in order to compare them to the ones obtained on C target.
 In fact, the two peaks observed in the $m_{cn}^2$ distribution of Fig.~\ref{f:dal}, that come from the horizontal
 bands in the Dalitz plot and are the signature of the sequential part of the decay, disappear completely with Pb.
 Instead, we observe the single, wider central peak characteristic of direct breakup, together with a much
 stronger effect of the $n$-$n$ FSI at low $m_{nn}^2$, clearly above the phase-space distribution (dotted line).
 We note, however, that the decay energy spectrum (inset) is comparable to the one in Fig.~\ref{f:Ecn},
 suggesting that we are populating a similar continuum region in $^8$He.
 
 Therefore, if our analysis of the sequential decay on C target is well founded, and with the Pb target we have
 switched off the sequential branch, the data on Pb should correspond to the direct component present in the C set.
 For completeness, we have compared the Pb data to the simulations of direct breakup using $P(^7\rm{He})=0$
 and the spatial $n$-$n$ configuration previously obtained, $\rms=7.3$~fm (Fig.~\ref{f:mc2}).
 The very good agreement with the data suggests that the breakup into $^6$He+$n$+$n$ on Pb does populate
 states in the continuum of $^8$He in the same energy range and with similar spatial characteristics, but that
 decay mainly through the simultaneous emission of both neutrons. This different decay mode can be understood at
 these bombarding energies by the effect of the stronger Coulomb field of Pb \cite{Iek93,Nak06}, that by acting
 only on the core subsystem hinders the probability of core-$n$ resonances (here $^7$He) to be formed.

\begin{figure}[t]
\begin{center}
 \includegraphics[width=0.36\textwidth]{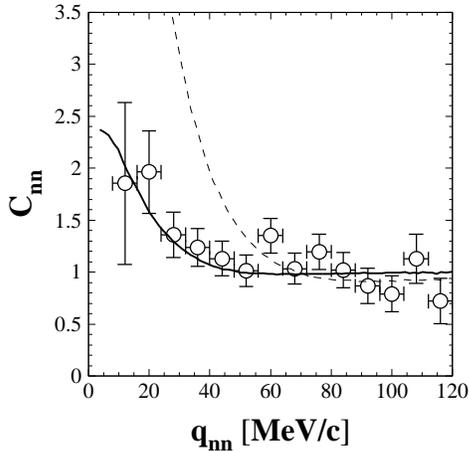}
 \caption{Two-neutron correlation function for the dissociation of $^8$He into $^6$He+$n$+$n$ on a C target.
 The dashed line corresponds to direct breakup with $\rms=7.3$~fm, and the solid line to an additional delay
 of $\tau=1400$~fm/$c$ for 70\% of events.} \label{f:Cnn}
\end{center}
\end{figure}

 Finally, we have built the experimental $C_{nn}(q_{nn})$, the observable used to parametrize the $n$-$n$ FSI
 in the model, for the dissociation on C. The experimental distribution was divided by a distribution obtained
 through event mixing, using the iterative technique described in Ref.~\cite{FMM00}. The resulting ratio is shown
 in Fig.~\ref{f:Cnn}. The dashed line corresponds to the analytical formula (\ref{e:Cnn0}) for
 $\rms=7.3$~fm that rises up to $C_{nn}(0)\approx11$, a value comparable to the ones measured in Ref.~\cite{FMM00}.
 If we use the more general formalism with $(\rms,\tau c)=(7.3,0)$~fm for 30\% of events and $(7.3,1400)$~fm for
 the other 70\%, we obtain the solid line. The agreement with the data is remarkably good, confirming that in this
 particular case the model does not only reproduce a general trend, but also the fine details of the $n$-$n$
 space-time signal.

\Header{Conclusions and outlook}
 We have measured the dissociation of $^8$He into $^6$He+$n$+$n$ on C and Pb targets. The analysis of triple
 correlations has given access, for the first time, to the spatial and temporal characteristics of the decay.
 The use of both targets has proven to be, at these energies, 
 an efficient way to switch on and off a delay between the emission of both neutrons, by selecting a mostly
 sequential (C) or a mainly direct (Pb) decay mode. The sequential channel is clearly identified through the
 population of the ground state of $^7$He, and represents a 70\% contribution in the case of the C target.

 The parallel analysis of the $n$-$n$ FSI in both target runs represents a very stringent test of this method
 as a femtometer {\it and\/} chronometer of $2n$ decays. In the C target case, it leads to a sequential decay from
 a Gaussian volume of $\rms=7.3\pm0.6$~fm, corresponding to states in the continuum of $^8$He a few MeV beyond
 the $^6$He+$n$+$n$ threshold, with a delay between the emission of both neutrons of $\tau=1400\pm400$~fm/$c$,
 consistent with the lifetime of $^7$He. On the other hand, the results for the Pb target clearly indicate a
 different mechanism favoring direct breakup, as expected for the stronger Coulomb field, that is well reproduced
 using the same $n$-$n$ volume without any delay.
 
 Regarding the spatial information, it would be interesting to compare this result with theoretical calculations
 of the $n$-$n$ distribution in the continuum of $^8$He, similar to those performed for $^6$He \cite{Dan04}.
 However, as noted earlier, $^8$He is better described as a five-body system. 
 In this respect, it would be interesting to consider a nucleus with a predominant three-body structure, like
 a heavier two-neutron halo, that at the same time possessed few, narrow states in the core-$n$ subsystem.
 For example, $^{17}$B exhibits a two-neutron halo \cite{Yam04} and the unbound subsystem $^{16}$B seems to have
 only one $^{15}$B-$n$ resonance extremely narrow, $\Gamma<100$~keV \cite{Lec09,Spy10}. That resonance, if
 populated through the sequential decay of $^{17}$B, would introduce a delay $\tau>2000$~fm/$c$ that should
 strongly hinder any $n$-$n$ correlation. We note that the breakup of $^{17}$B on C/Pb targets has been recently
 studied at RIKEN \cite{Orr12}, and that the analysis of $n$-$n$ correlations is in progress.
 
 The origin of the difference in the interpretation of these core-$n$-$n$ final states and the analogous
 core-$p$-$p$ ones is not clear yet. While the microscopic structure of the $2p$ emitter, a rather narrow state,
 seems to govern the $p$-$p$ distributions \cite{Ego12,Gri09,Gri09b}, the $n$-$n$ ones appear to be dominated
 by the effects of the $s$-wave $n$-$n$ FSI \cite{FMM00,FMM01,Rev18,Nor04,Gol05,Smi16,Spy12},
 as if the system went directly into the three-body continuum.
 In fact, in all the neutron works the systems were populated in a broad continuum of decay energies.
 Even if the $n$-$n$ scattering amplitude is much stronger than the $p$-$p$ one,
 unhindered by the Coulomb repulsion that almost cancels the latter,
 it would be interesting to study these correlations in a core-$n$-$n$ system with a well-defined energy,
 that could reveal the eventual breakdown of the model used here and/or the sensitivity to the microscopic
 structure of the $2n$-emitter state \cite{Gri18}. 

\section*{Acknowledgments}
 The support provided by the technical staff of LPC and the LISE crew is gratefully acknowledged, as are the
 efforts of the GANIL cyclotron operation team for providing the primary beam. We also wish to express our
 appreciation for the vital contributions made by our late colleague and friend Jean-Marc Gautier to all the
 CHARISSA+DEMON experiments undertaken by our collaboration.

\section*{Appendix: correlation formulas} \renewcommand\theequation{A\arabic{equation}}
 In the formalism of Ref.~\cite{Led82}, the correlation function for neutrons of four-momenta $p_i$
 emitted at a space-time relative distance $x=(\vec{r},t)$ has two terms, originating respectively from Fermi
 statistics and the $s$-wave FSI, averaged over the distribution of distances:
\begin{eqnarray}
 \lefteqn{C_{nn}(p_1,p_2) = 1 + \langle b_0\rangle + \langle b_i\rangle}	\\
 \langle b_0\rangle & = & -\frac{1}{2}\,\langle\cos(qx)\rangle	\\[-1mm]
 \langle b_i\rangle & = & \frac{1}{2}\,\Big\{ |f(k^\star)|^2\langle|\phi_{p_1p_2}(x)|^2\rangle	\nonumber \\
   & & +2\,\Re\left[f(k^\star)\langle\phi_{p_1p_2}(x)\cos(qx/2)\rangle\right]\Big\} \label{e:phi}
\end{eqnarray}
 The metric is such that $p^2=|\vec{p}|^2-p_0^2$, the superscript $^\star$ refers to the $2n$ center of mass,
 $q=p_1-p_2$ is the relative four-momentum, $k^\star=\sqrt{q^2}/2$ is the four-momentum of each neutron,
 and $f$ is their scattering amplitude:
\begin{equation}
 f(k^\star) = \left(-1/a_{nn}+k^{\star2}d_0/2-ik^\star\right)^{-1}
\end{equation}
 depending on the scattering length $a_{nn}$ and effective range $d_0$ (we use $-18.5$ and $2.8$~fm,
 respectively \cite{Wir95}). The $2n$ wave function is factorized assuming $r^\star\gtrsim d_0$
 as $f(k^\star)\phi_{p_1p_2}(x)$, with the exact form of $\phi_{p_1p_2}(x)$ given in Ref.~\cite{Led82}.
 However, note that the final expression of $C_{nn}$ (\ref{e:Bi}) does not depend on the form of $\phi_{p_1p_2}(x)$.
 
 If we now assume a spherically symmetric source $W$ and neglect its momentum dependence:
\begin{eqnarray}
 \langle b_0\rangle & = & -\frac{1}{2} \int W(x)\, \cos(qx)\, d^4x	\\[-1mm]
 \langle b_i\rangle & = & \int\! 2\pi r_T dr_T dr_L dt\ W(x)\, \Big\{ |f(k^\star)\phi_{p_1p_2}(x)|^2	\nonumber \\[-2mm]
   & & \hskip-8mm{+2\Re[f(k^\star)\phi_{p_1p_2}(x)] J_0\!\left(\frac{q_Tr_T}{2}\right)\cos\!\left(\!q_0\frac{r_L-vt}{2v}\right)\!\!\Big\}}
\end{eqnarray}
 with $L/T$ the directions parallel/perpendicular to the velocity $v$ of the pair.
 For a Gaussian source of the form $W(x)=\exp(-r^2/4r_0^2-t^2/4\tau_0^2)$ and small enough energies
 ($k^\star\ll m$), after integration over $t^\star$:
\begin{eqnarray}
 \langle b_0\rangle & = & -\frac{1}{2}\, \exp(-4k^{\star2}r_0^2-q_0^2\tau_0^2) \label{e:B0}	\\[-1mm]
 \langle b_i\rangle & = & \mbox{$\frac{1}{2\sqrt{\pi}r_0^2\gamma\rho}$} \int\! r_T dr_T dr^\star_L \exp(-r_T^2/4r_0^2-r^{\star2}_L/4\gamma^2\rho^2)	\nonumber \\[-.5mm]
   & & \hskip-9mm\times\!{\left\{\frac{|f|^2}{2r^{\star2}} + \Re\!\left[f\frac{\exp(ik^\star r^\star)}{r^\star}\right] J_0\!\left(\frac{q_Tr_T}{2}\right)\cos\!\left(\frac{q_0r^\star_L}{2\gamma v}\right)\right\}} \nonumber \\
   & & -(1/8\sqrt{\pi})|f|^2d_0/\gamma \rho r_0^2 \label{e:Bi}
\end{eqnarray}
 with $\rho=\sqrt{r_0^2+v^2\tau_0^2}$. The last term in (\ref{e:Bi}) is a first-order correction
 of the integration of the expression used for $\phi_{p_1p_2}(x)$ in the region $r^\star<d_0$.
 
 In the case of simultaneous emission and/or very small velocities ($\gamma\rho\approx r_0$)
 the final expression becomes analytical, with only one free parameter ($r_0$):
\begin{eqnarray}
 C_{nn}(q_{nn}) & = & 1 -\frac{1}{2} \exp(-q_{nn}^2r_0^2) +\frac{|f|^2}{4r_0^2}\left(1-\frac{d_0}{2\sqrt{\pi}r_0}\right)	\nonumber \\[-.5mm]
  & & +\frac{\Re f}{\sqrt{\pi}r_0} F_1(q_{nn}r_0) - \frac{\Im f}{2r_0} F_2(q_{nn}r_0) \label{e:Cnn0}
\end{eqnarray}
 with $F_1(z)=e^{-z^2}\!/z\int_0^ze^{x^2}dx$ and $F_2(z)=(1-e^{-z^2})/z$.
 Otherwise one should use (\ref{e:B0},\ref{e:Bi}), with two free parameters ($r_0,\tau_0$).
 From the source parametrization used, one obtains $\rms=\sqrt{6}r_0$ and $\tau=\sqrt{2}\tau_0$.

 Note that this model cannot be applied for $r_0\lesssim1$~fm ($\rms\lesssim2.5$~fm),
 since then the result is completely determined by the short-distance behavior of $\phi_{p_1p_2}(x)$
 in (\ref{e:phi}), sensitive to the form of the $n$-$n$ potential.

\end{document}